# Detection of ultra-weak laser pulses by free-running single-photon detectors: modeling dead time and dark counts effects

Hristina Georgieva[1], Alice Meda[2], Sebastian M. F. Raupach[1], Helmuth Hofer[1], Marco Gramegna[2], Ivo Pietro Degiovanni[2,3], Marco Genovese[2,3], Marco López[1], Stefan Kück[1]

[1] Physikalisch-Technische Bundesanstalt (PTB), Bundesallee 100, 38116 Braunschweig, Germany
[2] Istituto Nazionale di Ricerca Metrologica (INRIM), Strada delle Cacce 91, I-10135 Torino, Italy
[3] Istituto Nazionale di Fisica Nucleare (INFN), Sezione Torino, Via Giuria 1, I-10125 Torino, Italy

E-mail: stefan.kueck@ptb.de

**Abstract**

In quantum communication systems, the precise estimation of the detector´s response to the incoming light is necessary to avoid security breaches. The typical working regime uses a free-running single-photon avalanche diode in combination with attenuated laser pulses at telecom wavelength for encoding information. We demonstrate the validity of an analytical model for this regime which considers the effects of dark counts and dead time on the measured count rate. For the purpose of gaining a better understanding of these effects, the photon detections were separated from the dark counts via a software-induced gating mechanism. The model was verified by experimental data for mean photon numbers covering three orders of magnitude as well as for laser repetition frequencies below and above the inverse dead time. Consequently, our model would be of interest for predicting the detector response not only in the field of quantum communications, but also in any other quantum physics experiment where high detection rates are needed.

Keywords: single-photon detector, dead time, dark counts, laser pulses, quantum communication

Single-photon avalanche diodes (SPADs) are the most widespread commercial solution for single-photon technologies. In particular, quantum key distribution (QKD), a promising technique that allows two distant parties to share encryption keys with an unprecedented level of security, relies on SPADs for the photon detection [1-8]. As for all cryptographic systems, practical implementation of QKD requires a deep investigation and understanding of the employed devices to ensure correct operation [9-13]. For this reason, it is of utmost importance to employ theoretical models that precisely describe the operation of each component of the system. In fact, any deviation of a QKD device operation from the theoretical model can be exploited as a side channel or back door [14-17] by an eavesdropper to take the control of the process; in particular, several attacks related to the detection process are discussed in literature, such as the backflash based attacks [18-23] or the particularly critical group of detector-control attacks [24-32] including the famous blinding-attack [24]. For this reason, a big effort is dedicated to develop models and methods to characterize single QKD components. Such characterization techniques have to be validated and embedded in QKD standardization documents.

As already stated, the detectors are generally considered the most vulnerable part [18-32], so to guarantee security and avoid tailored attacks, the response of a SPAD to the incoming light in different regimes needs to be modelled and tested, taking into account its characteristic parameters (mainly quantum efficiency, dead time and dark counts) and a regime of operation as broad as possible. For example, the dead time is a critical parameter in QKD, since this effect limits the maximum count rate of such devices [33], in particular when the key transmission rates are increased [34,35] and security assumptions are violated [36]. Also, the number of dark counts should be correctly estimated in a QKD transmission, since their presence limits the achievable distance and leads to dead time effects, blinding the detector in the same way as a photon [37-41].

In most QKD systems, the asynchronous (free-running) regime of operation of SPADs is employed, while attenuated pulsed light is used for encoding quantum information. This particular regime was recently investigated [42], where a model to correct for the dead time and dark count effects was developed for a pilot study on the traceable measurement of the quantum efficiency of InGaAs/InP SPADs. The model was a useful approximation for the regime of that study, where a low mean photon number per pulse (up to 2.4) and a fixed repetition frequency were considered.

Here, we present and test an improved model for the count rate of the SPAD accounting for the dead time effects due to both weak laser pulses and dark counts. It is able to perfectly mimic the observed experimental behavior for an extremely wide variety of experimental conditions. We validate this model with an experiment investigating a range of mean photon numbers per pulse of up to 23 and repetition intervals of less than half the detector dead time. The agreement between the experimental data and the underlying model for these different modes of operation is excellent and allows the model to be used for predicting the response of a free-running detector to a pulsed source.

We begin with a short derivation of our model describing the SPAD response. Pulsed laser light with constant mean optical power in time has a number of photons per pulse that follows the Poisson distribution. Consequently, the probability $q$ for photon detection can be expressed as:

$$q = 1 - e^{-\eta\mu}. \qquad (1)$$

This probability is a function of the mean photon number per pulse $\mu$ and of the ideal constant detection efficiency $\eta$, which is not affected by any detector imperfections such as dark counts, dead time and afterpulsing. Next, we consider the effect of a fixed (non-extended) [33] dead time $D$ on the photon detection performed by a free-running SPAD detector. Figure 1(c) shows how the sensitivity of the SPAD detector changes in response to the incoming events in figure 1(b). Any absorbed photons within the "off" state of the detector will not produce an amplified detectable electronic output and will be lost. This effect can occur only if the dead time $D$ is larger than the time interval of $1/f$ between two consequent laser pulses ($fD > 1$). In particular, we expect that for any detection event there are on average $\text{Int}(fD)q$ lost events, where Int stands for the integer part, and $\text{Int}(fD)$ corresponds to the (deterministic) number of pulses falling in the dead time period after a detection event. Thus, the ideal detection rate would be $fq = fp_{\text{click}}[1 + \text{Int}(fD)q]$ with click probability $p_{\text{click}}$, or

$$p_{\text{click}} = \frac{q}{1+\text{Int}(fD)q}. \qquad (2)$$

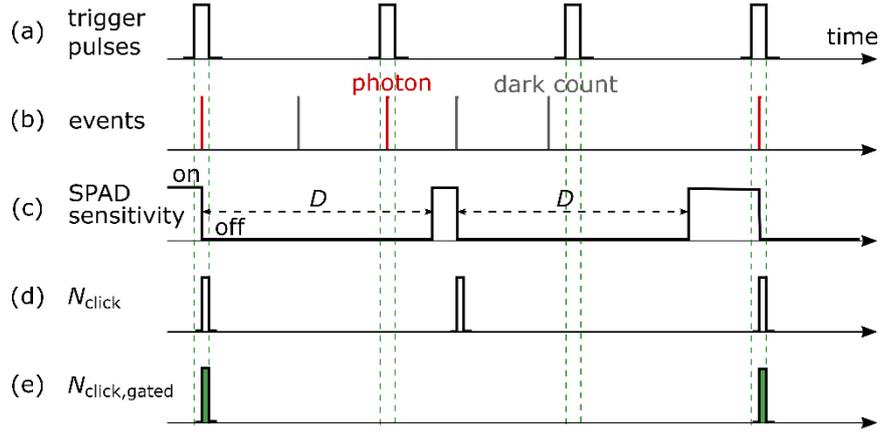

**Figure 1.** Schematic illustration of the response of a SPAD detector with a fixed dead time $D$ and filtering of the clicks correlated with laser pulses. (a) Trigger pulses with a period $1/f < D$. (b) Incoming events: photons (red) and dark counts (grey). Each laser pulse consists on average of $\mu$ photons, where $\mu$ is the mean photon number per pulse. (c) Changes of the SPAD sensitivity in response to the incoming events. After each detection the SPAD switches to the 'off' state for a time period $D$. (d) Electrical output of the SPAD detector: events within the dead time are lost. (e) Selection of detector clicks (green) which are correlated in time with trigger pulses and thus with high probability originate from real photon events.

At this point, the effect of dark counts must also be taken into account. The dark count rate is measured with the laser switched off and stray light completely blocked. The number of clicks from laser pulses is usually estimated by subtracting the measured dark count rate from the effective count rate. In the following, we develop a more comprehensive and detailed model showing the limit of that simple approach.

In the presence of detection counts originating from laser pulses we have to account for the fact that the associated dead time reduces the available measurement time during which the detector may detect dark counts (see figure 1(a)-(c)). Since the expected detection rate for laser pulses is approximately $f\,p_{\text{click}}$, the detector during a time period $T$ will not be blocked by dead time effects only for a reduced time period of $T - T\,f\,p_{\text{click}}\,D$. This reduction in effective detection time leads to an expected reduction of the dark count rate by $(1 - f\,p_{\text{click}}\,D)$, i.e.

$$N_{\text{dark}} = N_{\text{dark,exp}}\,(1 - f\,p_{\text{click}}\,D). \tag{3}$$

This equation does not include the effect of afterpulsing, which can be neglected for a sufficiently high dead time. The total rate of the measured clicks accounts for the contribution of the rate of detection events due to laser pulses and dark counts, i.e.

$$N_{\text{click}} = f\,p_{\text{click}}\,p_{\text{no dc}} + N_{\text{dark}}, \tag{4}$$

where $p_{\text{no dc}} = \exp(-N_{\text{dark,exp}}D)$ approximately (under the condition $N_{\text{dark,exp}}D \ll 1$) corresponds to the probability of the absence of dark counts in a time interval $D$ before the laser pulse detection event. The final equation for the SPAD count rate is therefore:

$$N_{\text{click}} = f \frac{1-e^{-\eta\mu}}{1+\text{Int}(fD)(1-e^{-\eta\mu})} e^{-N_{\text{dark,exp}}D} +$$
$$+ N_{\text{dark,exp}}\left(1 - \frac{1-e^{-\eta\mu}}{1+\text{Int}(fD)(1-e^{-\eta\mu})}fD\right). \tag{5}$$

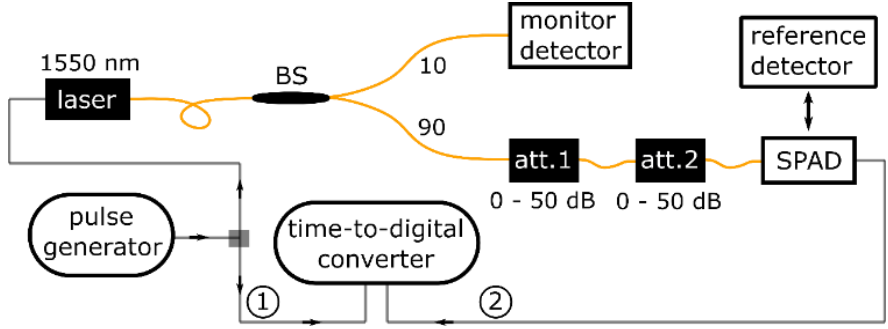

**Figure 2.** Sketch of the experimental setup for count rate measurements with a free-running SPAD under pulsed laser radiation with variable repetition frequency and optical power. A monitor detector connected to a beam splitter (BS) tracks any changes of the optical laser power. Two calibrated variable attenuators (att. 1 and att. 2) reduce the optical power by several orders of magnitude so that it lies below the saturation value of the SPAD. The electrical signals from the pulse generator and from the SPAD are connected to channels 1 and 2 of a time-to-digital converter for filtering of the clicks correlated with laser pulses.

The setup for count rate measurements with variable optical powers and pulse frequencies is presented in a schematic drawing in figure 2. A laser (IDQ, id300) with an emission wavelength of 1550.5 nm is triggered by a pulse generator (Keysight, 33600A) with a variable repetition frequency. The laser light, which is monitored in intensity, passes through two identical variable attenuators (Agilent, 81571A) and reaches the active area of an InGaAs/InP SPAD (IDQ, id220-FR-SMF). The arrival times of the electrical pulses from the pulse generator on channel 1 (figure 1(a)) and from the SPAD detector on channel 2 (figure 1(d)) are recorded by a time-to-digital converter (Swabian Instruments, Time Tagger 20). It should be noted that both signals have been synchronized by choosing an appropriate delay compensation on channel 1 to account for the difference in cable lengths. All measurements are performed in free-running mode. As can be seen in figure 1(e), post-processing of the time tags from the SPAD has been carried out to select events which coincide with the arrival of a trigger pulse within a small time interval of 3 ns. It is selected to be significantly smaller than the inverse repetition frequency, but larger than the timing jitter of the SPAD detector, which is equal to 0.52 ns. In this way, one can distinguish between "true" (photons) and "false" (dark counts) detection events, because the detection of a photon is always preceded by the emission of a laser pulse, whereas dark counts are uncorrelated in time.

We are interested in the number of events per time interval. For this purpose, the time-to-digital converter was operated in counter mode to measure the click rate $N_{\text{click}}$ of the SPAD, as well as the rate of clicks within the software-induced gate window $N_{\text{click,gated}}$ (figure 1(e)). They have been determined for different light intensities and pulse frequencies by taking the mean value of 100 measurements, each with an integration time of one second.

The mean photon number per pulse $\mu$ has been determined from two additional measurements, for which the SPAD detector has been replaced by a low-noise reference analog detector (Hamamatsu, G8605). First, the reference detector yields a traceable frequency-dependent value of the optical power at zero attenuation: $P_0(f) = I_0(f)/s(\lambda)$, where $I_0$ is the corresponding photocurrent and $s$ is the calibrated responsivity of the InGaAs reference diode for the same laser wavelength $\lambda$. Second, the attenuation factors $k^I$ and $k^{II}$ from the first and second attenuator have been calibrated according to the method presented in reference [43]. They have been corrected for changes in the laser power with the help of the monitor detector. The mean photon number is then given by:

$$\mu = \frac{\lambda I_0 k^I k^{II}(1+c_{\text{lin}})}{h c f s}, \tag{6}$$

where $f$ is the repetition frequency, $h$ and $c$ are physical constants (Planck constant and speed of light) and $c_{lin}$ is a linearity correction factor of the reference diode responsivity.

In the following, the model for $N_{click}$ (equation (5)) is compared with experimental data. It is important to note that our model does not allow for any degrees of freedom and therefore presents direct dependencies between the measured count rate and three SPAD properties: detection efficiency $\eta = 0.1$, dark count rate $N_{dark,exp} = 805.2$ counts/s and dead time $D = 20.31$ µs, as well as two variable properties of the laser source: repetition frequency $f$ and mean photon number per pulse $\mu$. For the selected combination of dead time and detection efficiency, the afterpulsing probability has been determined to $(0.05 \pm 0.01)\,\%$. Therefore, afterpulsing has a negligible effect on the data within the stated uncertainty.

One advantage of our setup is that it enables the experimental determination of the gated count rate $N_{click,gated}$. Due to the small gating time of only 3 ns compared to the smallest interval of 5.9 µs between two consecutive laser pulses, $N_{click,gated}$ is approximately equal to the photon count rate described by the first term in equation (5). Therefore, we are able to indirectly determine the dark count rate, since $N_{dark} = N_{click} - N_{click,gated}$.

For the first data set, the mean photon number has been varied for three pulse frequencies: 30 kHz, 80 kHz, 130 kHz, which have been selected to cover three different cases: $\text{Int}(fD) = 0, 1, 2$. The results are presented in figure 3. Our model (red curves) gives a very good estimation of the measured count rate in all three cases. As can be seen in figure 3(a), a change in the repetition frequency affects not only the slope at low $\mu$, but also the saturation value that is reached at high laser power. For the first case, where the repetition frequency lies below the inverse dead time of 49.24 kHz (black diamonds), the nonlinear saturation effect is caused only by the photon statistics of the laser source. Since the SPAD detector can only produce one 'click' for one or more photons per pulse, the maximal count rate is given by the repetition frequency $f$. At higher frequencies, the dead time has a strong influence on the saturation behavior and the count rate is limited by $f/[\text{Int}(fD) + 1]$. It should also be noted that equation (5) is a function of the product $\eta\mu$, meaning that if the detector has a higher detection efficiency, saturation will be reached at a lower mean photon number.

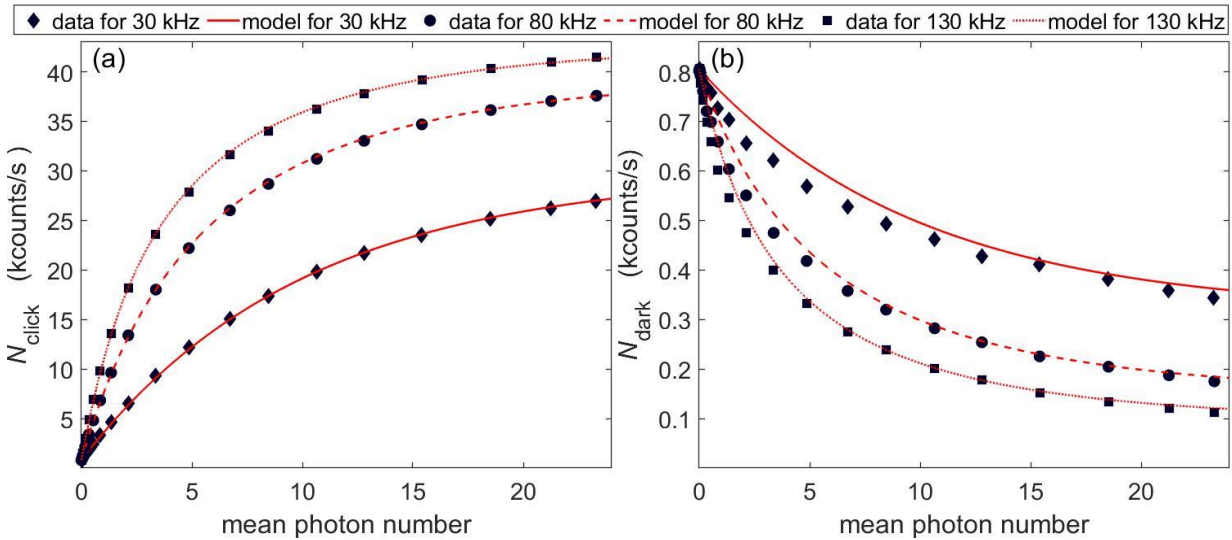

**Figure 3.** (a) $N_{click}$ and (b) $N_{dark}$ as a function of the mean photon number for three different repetition frequencies: 30 kHz (diamonds), 80 kHz (circles) and 130 kHz (squares). The red curves represent (a) the model for $N_{click}$ from equation (5) and (b) the model for $N_{dark}$ from equation (3). For error bars see figure 5.

The model is also able to correctly describe the decrease of the dark count rate for an increasing mean photon number (figure 3(b)). As $\mu$ approaches zero, one obtains $N_{\text{dark,exp}}$: the dark count rate that can be directly measured in the absence of light. There is a discrepancy between model and experiment of less than 55 counts per second for mean photon numbers between 1 and 5. This small deviation lies within the expanded measurement uncertainty. Possible systematic effects are subject to further investigations.

Next, we focus on the frequency dependence of $N_{\text{click}}$ and $N_{\text{dark}}$ for three selected mean photon numbers: 0.19, 1.5 and 20.5. Equation (5) predicts a discontinuous change of the count rate for frequencies corresponding to a multiple integer of the inverse dead time. As can be seen in figure 4, the height and position of these jumps are described quite well by our model. There is a change in slope depending on the frequency region. As expected, the local maxima in $N_{\text{click}}$ correspond to local minima in $N_{\text{dark}}$, because the total blind time of the detector due to photon detection, each triggering a dead time of 20.31 µs, increases at higher count rates. The height of the jumps gets negligibly small for $\mu = 0.19$ and the curve seems to have a constant slope. This slope is correctly predicted for $N_{\text{click}}$, whereas it is slightly underestimated for $\mu = 20.5$ in figure 4(b). One can conclude that for values of $\eta\mu$ approaching zero, there is a linear dependence between count rate and repetition frequency. However, in the regime of high optical power and/or high detection efficiencies, the specific sawtooth shape of $N_{\text{click}}$ and $N_{\text{dark}}$ dominates. In this regime, it seems advisable to avoid measurements at frequencies that are an exact multiple of the inverse dead time, because the detector response cannot be accurately predicted.

Figure 5 shows a comparison of three different models for the dark count rate. Model A is given by the second term in equation (5):

$$N_{\text{dark}} = N_{\text{dark,exp}} \left(1 - \frac{1-e^{-\eta\mu}}{1+\text{Int}(fD)\cdot(1-e^{-\eta\mu})} fD\right). \tag{7}$$

Model B is taken from reference [42]. This model also includes a dead time correction for the dark counts as a function of the mean photon number and laser frequency:

$$N_{\text{dark}} = \frac{N_{\text{dark,exp}}}{1 + [f(1-e^{-\eta\mu})+N_{\text{dark,exp}}]D}. \tag{8}$$

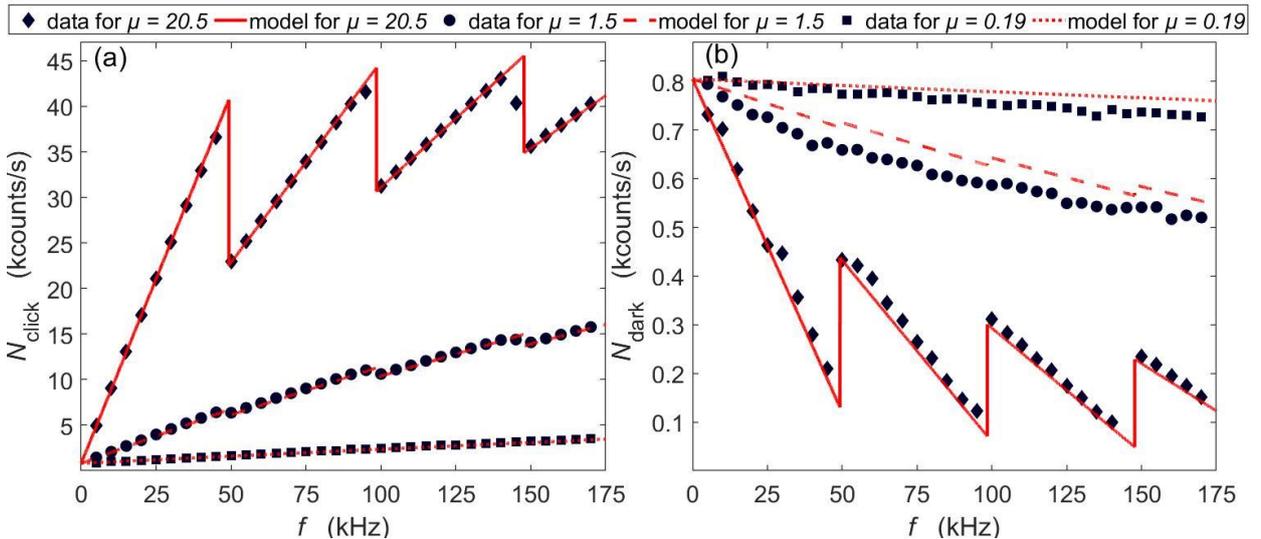

**Figure 4.** (a) $N_{\text{click}}$ and (b) $N_{\text{dark}}$ as a function of the laser repetition frequency for three different mean photon numbers: 0.19 (squares), 1.5 (circles) and 20.5 (diamonds). The red curves represent (a) the model for $N_{\text{click}}$ from equation (5) and (b) the model for $N_{\text{dark}}$ from equation (3). For error bars see figure 5.

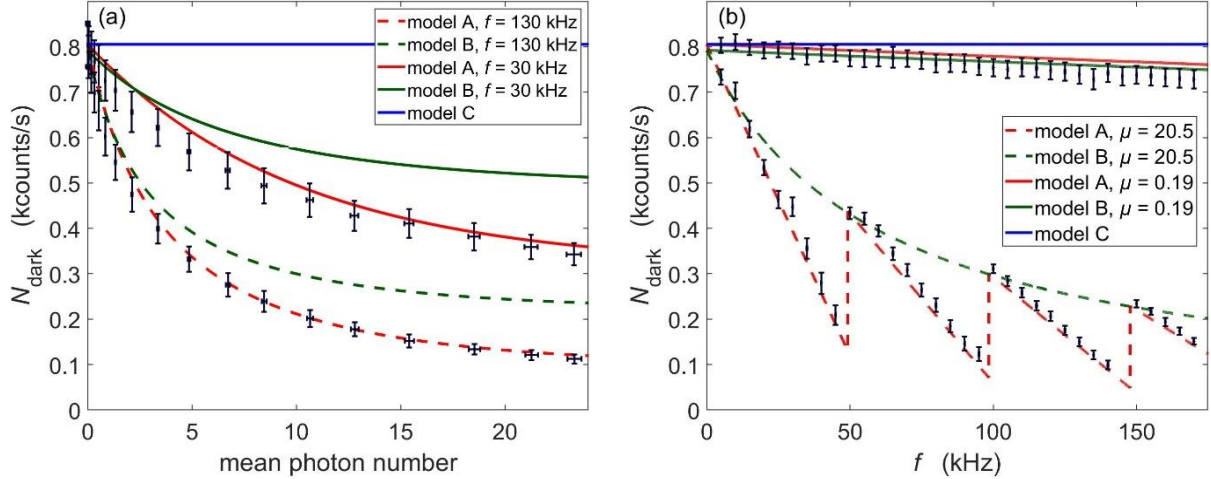

**Figure 5.** Comparison of three models for $N_{\text{dark}}$. Model A (equation (7)) is from this letter, model B (equation (8)) is from reference [42]. Model C (equation (9)) is the usual assumption of a constant dark count rate. The black error bars in x and y represent the expanded measurement uncertainty (k = 2) (a) The repetition frequency is fixed ($f$ = 30 kHz and $f$ = 130 kHz) and the mean photon number is varied. (b) The mean photon number is fixed ($\mu = 0.19$ and $\mu = 20.5$) and the repetition frequency is varied.

Last but not least, model C contains the simple assumption of a constant dark count rate:

$$N_{\text{dark}} = N_{\text{dark,exp}}. \tag{9}$$

This assumption has a widespread usage in many applications including count rate measurements with SPAD detectors of any kind [39].

Model A gives by far the best description of the data in figure 5. Model B is unable to follow the sawtooth shape of the frequency dependence, but it describes well the dark counts at repetition frequencies slightly above a multiple integer of the inverse dead time (figure 5(b)). Moreover, the discrepancy between models A and B in figure 5 begins to vanish for $\eta\mu \to 0$. Consequently, model B gives a good approximation only for low mean photon numbers, where the frequency dependence is approximately linear.

From figure 5 it becomes clear that the assumption of model C of a constant dark count rate is not valid in general. Even for $\mu = 0.19$ (figure 5(b)), there is a visible decrease of the dark count rate with increasing frequency. As a rule of thumb, the smaller the product $N_{\text{dark,exp}}D$, the lower the dead time correction on the dark counts. For specific experimental conditions, one can use equation (7) to calculate the magnitude of this correction.

In this letter, we presented a model of the response of a free-running SPAD detector for pulsed light covering even pulse periods considerably shorter than the dead time. The model includes three detector parameters: dead time, dark counts and detection efficiency. We demonstrated an experimental method for separating photon detections from dark counts via a software-induced gating mechanism, which has been proven valuable for gaining a better understanding of the influence of dead time. Our model gives a much better description of the effective dark counts compared to previous models. Its validity has been verified through direct comparison with experimental data.

The presented model in principle should give a good estimation of the count rate for an arbitrary laser repetition frequency, so that it could be implemented in QKD-based applications aiming for high data rates. Further work should include extending the model to consider the effect of afterpulsing.

The model will be useful for testing the expected response of a free-running detector illuminated by a pulsed source in QKD devices to detect any deviation from the ideal behavior that can be exploited by an eavesdropper to gain information about the system.

**Acknowledgements**

The work reported on this letter was funded by the project EMPIR 19NRM06 METISQ. This project received funding from the EMPIR program co-financed by the Participating States and from the European Union Horizon 2020 research and innovation program. This research was supported by the following projects: project "Piemonte Quantum Enabling Technologies" (PiQuET), funded by the Piemonte Region within the "Infra-P" scheme (POR-FESR 2014-2020 program of the European Union); NATO under the SPS program, project "Analysis, design and implementation of an end-to-end 400 km QKD link" (grant SPS G5263). We gratefully acknowledge the support of the Braunschweig International Graduate School of Metrology B-IGSM and the DFG Research Training Group 1952 Metrology for Complex Nanosystems. This work was also supported by the Deutsche Forschungsgemeinschaft (DFG, German Research Foundation) under Germany's Excellence Strategy – EXC-2123 QuantumFrontiers - 390837967.

**Disclaimer**

The name of any manufacturer or supplier by PTB or INRIM in this scientific journal shall not be taken to be PTB's or INRIM's endorsement of specific samples of products of the said manufacturer; or recommendation of the said supplier. Furthermore, PTB and INRIM cannot be held responsible for the use of, or inability to use, any products mentioned herein that have been used by PTB and INRIM.

**Data Availability**

The data that support the findings of this study are openly available in ZENODO at http://doi.org/10.5281/zenodo.4483256.